\documentclass[aps,pra,twocolumn]{revtex4-1}
\usepackage{color,amsthm,amsmath,amsfonts,graphicx,bm,epsfig,float,siunitx,multirow,xr,enumerate,epstopdf,soul}
\usepackage[breaklinks=true,colorlinks=true,linkcolor=blue,urlcolor=blue,citecolor=blue]{hyperref}
\usepackage[T1]{fontenc}
\usepackage{physics,verbatim,ulem,mathrsfs}%caption
\usepackage[toc,page,title,titletoc,header]{appendix}

\begin{document}

\title{Heisenberg-limited spin squeezing in coupled spin systems}
\author{Long-Gang Huang$^{1}$}
\thanks{These authors contributed equally to this work.}
\author{Xuanchen Zhang$^{1}$}
\thanks{These authors contributed equally to this work.}
\author{Yanzhen Wang$^{1}$}
\thanks{These authors contributed equally to this work.}
\author{Zhenxing Hua$^{1}$}
\author{Yuanjiang Tang$^{1}$}
\author{Yong-Chun Liu$^{1, 2}$}
\email{ycliu@tsinghua.edu.cn}
\affiliation{$^{1}$State Key Laboratory of Low-Dimensional Quantum Physics, Department of
	Physics, Tsinghua University, Beijing 100084, P. R. China}
\affiliation{$^{2}$Frontier Science Center for Quantum Information, Beijing 100084, P.R.
	China}
\date{\today}

\begin{abstract}
Spin squeezing plays a crucial role in quantum metrology and quantum
information science. Its generation is the prerequisite for further
applications but still faces an enormous challenge since the existing
physical systems rarely contain the required squeezing interactions. Here we
propose a universal scheme to generate spin
squeezing in coupled spin models with collective spin-spin interactions, which commonly exist in various systems.
Our scheme can transform the coupled spin interactions into squeezing interactions, and reach the extreme squeezing with Heisenberg-limited measurement precision scaling as $1/N$ for $N$ particles. Only constant and
continuous driving fields are required, which is accessible to a series of
current realistic experiments. This work greatly enriches the variety of
systems that can generate the Heisenberg-limited spin squeezing, with broad
applications in quantum precision measurement.
\end{abstract}

\maketitle

\section{introduction}

Squeezed spin states (SSSs) are entangled quantum states of collective spins
with reduced quantum fluctuations in one spin component perpendicular to the
mean spin direction, due to the quantum correlation between spins \cite%
{Kitagawa1993,Wineland1992,Wineland1994,Ma2011}. The reduced quantum
fluctuations allow them to surpass the so-called standard quantum limit
(SQL) with measurement precision scales as $\propto 1/N^{1/2}$ for $N$
particles \cite{Bohnet2014,Hosten2016science,Hosten2016,Luo2017,Bao2020},
which is permitted by the coherent spin states (CSSs). Thereby, the SSSs are
the key resources in the field of quantum metrology, which have significant
applications in high-precision measurements \cite%
{Wineland1992,Wineland1994,Andre2004,Meiser2008,Polzik2008,Gross2010,Riedel2010,Leroux2010,Sewell2012,Cox2016,Pezze2018,Kaubruegger2019,Pedrozo-Penafiel2020,Szigeti2020}%
. Due to their close relationship with quantum correlation, they also serve
as a significant witness to reveal the many-particle entanglement, which
have attracted extensive research interests during the past few decades \cite%
{Sorensen2001nature,Sorensen2001prl,Wang2003,Korbicz2005,Korbicz2006,Esteve2008,Toth2009,Hyllus2012,Toth2014,Bohnet2016,Ren2021,Feng2021}%
. Since their preparation is the prerequisite for further applications, many
efforts have been made to produce the SSSs, mainly in two categories of
platforms: atom-light interactions \cite%
{Hald1999,Takeuchi2005,Schleier-Smith2010,Hammerer2010,Leroux2010,Yu2014,Zhang2015,Zhang2017,Qin2020,Groszkowski2020}
and nonlinear atom-atom interactions, e.g., Bose-Einstein condensates (BECs)
\cite{Sorensen2001nature,Orzel2001,Esteve2008,Gross2010,Riedel2010,Fadel2018}%
.

Among the category of atom-light interaction platform, one method of
generating spin squeezing is by transferring the squeezing from squeezed
light to spin system \cite{Kuzmich1997,Hald1999,Vernac2000,Fleischhauer2002}%
. It is straightforward but limited by the transfer efficiency and the
performance of light squeezing. Besides, some proposals use  photon-mediated spin-spin interactions generated in an optical cavity to obtain SSSs \cite{Matthew2018,Lewis2018}, in which case superradiance is the main restriction on the achievable squeezing. Quantum nondemolition (QND) measurement is
another experimentally feasible way to generate spin squeezing \cite%
{Kuzmich2000,Appel2009,Chen2011PRL,Hosten2016,Rossi2020,Kritsotakis2021}%
, but the acquired squeezing is not deterministic and therefore strongly depends on the
performance of the photodetector \cite{Kuzmich1998,Pezze2018}. For the nonlinear atom-atom interaction platform, two well-known
mechanisms, i.e., one-axis twisting (OAT) and two-axis twisting (TAT) can
deterministically generate the spin squeezing \cite%
{Kitagawa1993,Ma2011,Liu2011,Zhang2014,Huang2015,Shen2013,Hu2017,Chen2019}.
Interparticle interactions in BEC lead to OAT dynamics in certain circumstances, which has been realized in several experiments to create metrologically useful squeezing \cite{Gross2010,Riedel2010}. Studies in trapped ions \cite{Bohnet2016,Figgatt2019,Lu2019} and superconducting qubits \cite{Song2017,Song2019,Xueaba2020} also witness OAT interaction and use it for entanglement generation. Though the OAT squeezing has been experimentally demonstrated in these systems, the squeezing degree only scales as $\propto 1/N^{2/3}$, which is still
far from the Heisenberg-limited measurement precision. In contrast, the TAT squeezing can provide a fascinating squeezing degree scaling as $\propto 1/N$, which can reach the Heisenberg-limit measurement precision, but its generation remains
a great challenge since the interaction form is not found naturally in
current realistic physical systems \cite%
{Helmerson2001,Liu2011,Shen2013,Zhang2014,Huang2015,Zhang2017,Hu2017,Wang2017,Zhang2017,Borregaard2017,Chen2019,Macri2020}%
. For certain existing interactions with weak or without squeezing ability, utilization of pulse sequences is shown to possibly induce engineered OAT or TAT Hamiltonian \cite{Cappellaro2009,Ben2020,Choi2020,Zhou2020,Huang2021}.
Several other theoretical schemes are devoted to transforming the OAT interaction into a TAT type to approach the ultimate Heisenberg limit \cite{Liu2011,Shen2013,Zhang2014,Huang2015,Chen2019}. However, either special experimental systems or complicated designs are required in these proposals. Therefore, it is essential to explore feasible schemes capable of realizing Heisenberg-limited squeezing with commonly-existed systems and easily-implementable designs.

Here we propose a universal scheme to produce Heisenberg-limited spin
squeezing in generic coupled spin systems with collective spin-spin interactions by continuous drivings. An
effective OAT interaction can be induced by simply applying a constant
direct-current (DC) field, leading to strong spin squeezing. We can further
obtain the TAT spin squeezing with an additional continuous
alternating-current (AC) field, and the Heisenberg-limited measurement
precision is ultimately achieved. Unlike schemes such as those reported in Ref. \cite{Muessel2015} and Ref. \cite{Haine2020}, which require an intrinsic spin-squeezing interaction that can be
enhanced through drives, our scheme focuses on generating spin squeezing from a collective interaction that does not intrinsically result in spin squeezing. As a result, our scheme largely enriches the variety of systems that can realize or enhance spin squeezing. Meanwhile, unlike previous studies, our approach only needs a constant field to generate
effective OAT interaction and an additional continuous driving to generate
effective TAT interaction, which is favorable for experimental
implementation.

\section{system model}

A broad category of coupled spin systems can be universally described by the collective
interaction Hamiltonian
\begin{equation}
	H_{\mathrm{int}}=\sum\nolimits_{\mu }g_{\mu }S_{\mu }J_{\mu },  \label{H}
\end{equation}%
where $g_{\mu }(\mu =x,y,z)$ denotes the coupling strength between the two
subsystems for different spin components, described by the collective spins
(or pseudospins) \textbf{S} and \textbf{J}, respectively. The operators are defined as $S_{\mu }=\sum_{k=1}^{N_{s}}\sigma _{S,\mu
}^{(k)}/2$ and $J_{\mu }=\sum_{k=1}^{N_{j}}\sigma _{J,\mu }^{(k)}/2$,
denoting the collective spin components, with $\sigma _{S,\mu }^{(k)}$ and $%
\sigma _{J,\mu }^{(k)}$ being the corresponding Pauli matrices for the $k$%
-th spin-$1/2$ (or two-level) particle. They can also describe the Stokes
operators of light, which are related to the differences between the number
operators of the photons polarized in different orthogonal bases \cite%
{Ma2011,Pezze2018}. The operators satisfy the SU(2) angular momentum
commutation relations $[S_{i},S_{j}]=i\varepsilon _{ijk}S_{k}$ and $%
[J_{i},J_{j}]=i\varepsilon _{ijk}J_{k}$, where $\varepsilon _{ijk}$ ($%
i,j,k=x,y,z$) is the Levi-Civita symbol. The above model can be used to
describe the atom-light interaction system \cite{Kuzmich1998,Bao2020},
spin-exchange interaction system, dipole-dipole interaction system, etc. The corresponding typical
interaction Hamiltonians are $%
H_{1}=gS_{z}J_{z},H_{2}=g(S_{x}J_{x}+S_{y}J_{y}+S_{z}J_{z})$ and $%
H_{3}=g(S_{x}J_{x}+S_{y}J_{y}-2S_{z}J_{z})$, as illustrated in Fig.~\ref%
{fig:1}(a), (b) and (c), respectively.

\begin{figure}[t]
	\includegraphics[width=1\columnwidth]{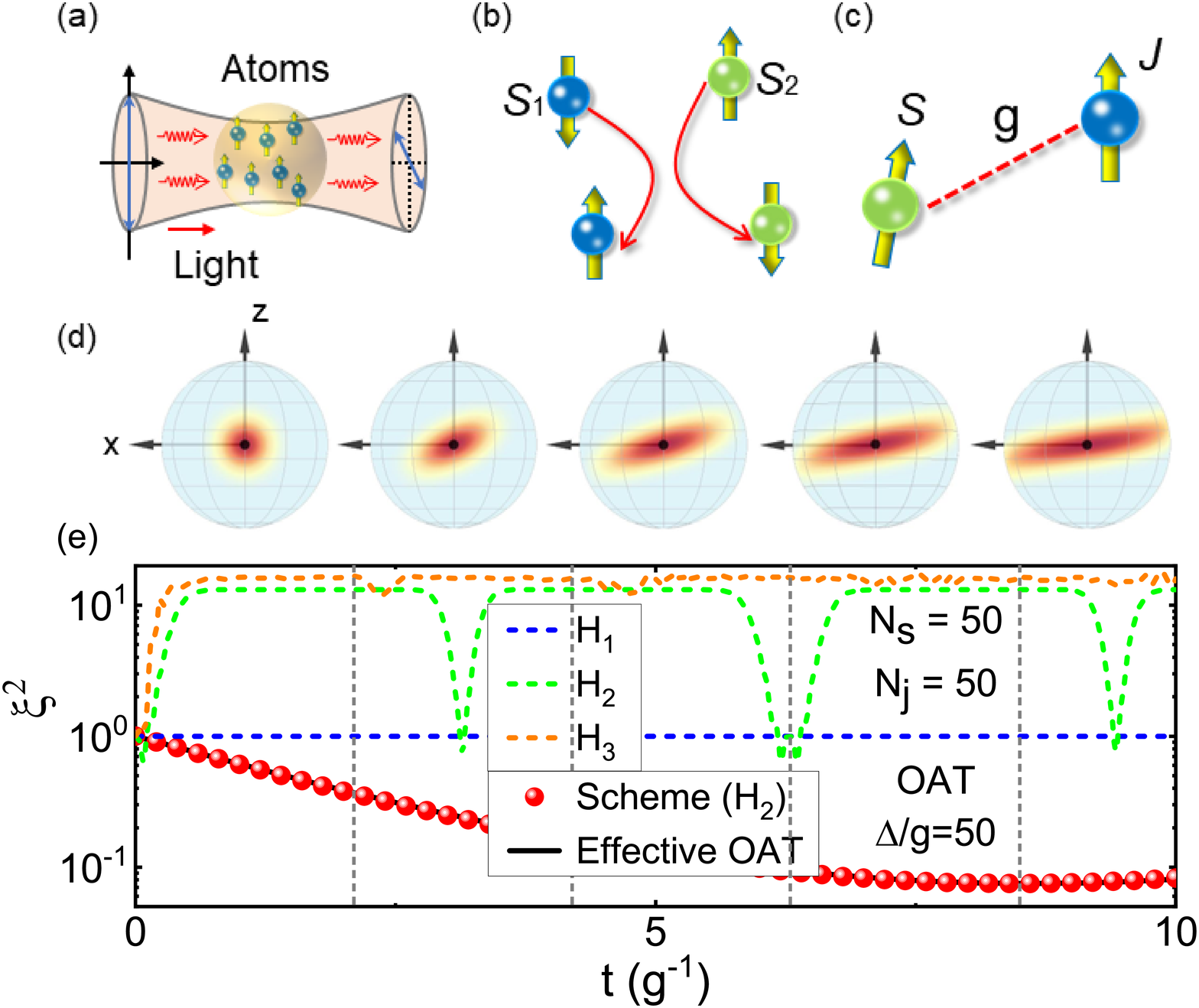}
	\caption{Schematic diagram of coupled spin systems and spin squeezing via
		constant drivings. (a) Schematic diagram of atom-light interaction system
		with Faraday magneto-optic rotation. (b) Schematic diagram of spin-exchange
		interaction between two spins. (c) Schematic diagram of
		dipole-dipole interaction between two spins. (d)
		Evolution of the quantum state for spin $S$ at the time instants denoted by
		the vertical gray dashed lines in (e), represented by the Husimi Q function
		on the generalized Bloch spheres. (e) The blue dashed line, green and orange dashed
		curves respectively present the free evolution of the squeezing parameters $%
		\protect\xi ^{2}$ for spin $S$ under the Hamiltonians $H_{1}$, $H_{2}$ and $%
		H_{3}$ as defined in the main text. The red solid balls denote the results
		of our scheme with $H_2$ under constant drivings, compared
		with the effective OAT interaction (\protect\ref{HOATeff}) (black solid
		curve). The vertical gray dashed lines mark time instants of $t=t_{\mathrm{%
				min}}/4,t_{\mathrm{min}}/2,3t_{\mathrm{min}}/4$ and $t_{\mathrm{min}}$, with
		$t_{\mathrm{min}}$ being the optimal squeezing time of OAT. The parameters
		are $N_{s}=50$, $N_{j}=50$ and $\Delta /g=50$. The initial state is product
		of coherent spin states polarized along $y$ axis for spin $S$ and $z$ axis
		for spin $J$.}
	\label{fig:1}
\end{figure}

Considering the practical applications, we focus on the spin squeezing of
one subsystem, e.g., spin $S$. The interaction Hamiltonian (\ref{H}) does
not contain the intra-species nonlinear interaction form like $S_{\mu }^{2}$%
, thus not being able to generate the OAT interactions. We
apply constant DC driving fields on both subsystems, generally described by
\begin{equation}
	H_{\mathrm{dr\mathrm{i}v}}^{\mathrm{DC}}=\Omega J_{z}+\Omega ^{\prime }S_{z}.
	\label{Hrot}
\end{equation}%
Here $\Omega $ and $\Omega ^{\prime }$ denote the magnitudes of homogeneous
fields along $z$ axis applied on spins $S$ and $J$, respectively. Thereby,
the total Hamiltonian becomes $H_{\mathrm{tot}}=H_{\mathrm{int}}+H_{\mathrm{%
		driv}}^{\mathrm{DC}}$, which will be demonstrated to be equivalent to an
effective OAT form $\propto S_{z}^{2}$ in the following.

To be specific, the mechanism of inducing OAT spin squeezing is analogous to
that of electron-phonon interaction in condensed matter physics, in which
the effective electron-electron interaction is mediated by the phonons
(lattice vibrations). In our scheme, the spin $J$ acts as an intermediary
(analogous to the role of phonons) to induce the intra-species interaction in the
spin $S$ (analogous to the role of electrons). The effective intra-species
interaction in spin $S$ can be derived by performing the Fr\"ohlich-Nakajima
transformation (FNT) \cite{frohlich1950theory,Nakajima1955} with $U=e^{S}$
on the total Hamiltonian of the coupled spin system as
\begin{align}
	H_{\mathrm{eff}}& ={e^{-S}}H_{\mathrm{tot}}{e^{S}}=H_{\mathrm{driv}}^{%
		\mathrm{DC}}+(H_{\mathrm{int}}+[H_{\mathrm{driv}}^{\mathrm{DC}},S])  \notag
	\\
	& +\frac{1}{2}[(H_{\mathrm{int}}+[H_{\mathrm{driv}}^{\mathrm{DC}},S]),S]+%
	\frac{1}{2}[H_{\mathrm{int}},S]+...,  \label{FNT1}
\end{align}%
Choosing an appropriate $S=-i\theta _{\mathrm{xy}}S_{x}J_{y}-i\theta _{%
	\mathrm{yx}}S_{y}J_{x}$ with $\theta _{\mathrm{xy}}$ and $\theta _{\mathrm{yx%
}}$ being undetermined coefficients, so that the first order term $H_{%
	\mathrm{int}}+[H_{\mathrm{driv}}^{\mathrm{DC}},S]$ almost vanishes, the Hamiltonian
is simplified as $H_{\mathrm{eff}}\simeq H_{\mathrm{driv}}^{\mathrm{DC}}+[H_{%
	\mathrm{int}},S]/2$. The commutator $[H_{\mathrm{int}},S]$ contains the quadratic terms $S_{\mu
}^{2}$, thus being able to generate the OAT spin squeezing (see Appendix \ref{App1} for detailed derivations).

Furthermore, the initial state of spin $J$ is chosen to be a CSS polarized along $z$ axis, i.e., the eigenstate of $J_{z}$ with
eigenvalue $N_{j}/2$, where $N_{j}$ is the corresponding particle number.
During short time evolution, the operators describing the spin $J$ can be
approximately replaced by their expectation values in the initial
state. As a
result, the Hamiltonian becomes
\begin{equation}
	H_{\mathrm{eff}}=fS_{z}+pS_{x}^{2}+qS_{y}^{2},  \label{Hlmgeff}
\end{equation}%
where $f,p,q$ are functions of $\Omega $ and $\Omega ^{\prime }$
(see Appendix \ref{App1}).
Appropriate combination of $\Omega $ and $\Omega ^{\prime }$ can make $f=0$,
eliminating the linear term in Eq.~(\ref{Hlmgeff}). When $p=q$ (the case is
similar when one of $p$ and $q$ is $0$), Eq.~(\ref{Hlmgeff}) is reduced to a
pure OAT Hamiltonian
\begin{equation}
	H_{\mathrm{eff}}^{\mathrm{OAT}}=\chi _{\mathrm{eff}}S_{z}^{2},
	\label{HOATeff}
\end{equation}%
with the effective nonlinear interaction strength $\chi _{\mathrm{eff}%
}=-g^{2}/(2\Delta )$. Here $\Delta =2(\Omega -\Omega ^{\prime })/(N_{j})$ is a parameter
characterizing the difference between the magnitudes of two external fields.
The condition $p=q$ can be satisfied when $g_{x}=g_{y}\equiv g$. Note that the sign and magnitude of the interaction strength $\chi_\mathrm{eff}$ can be easily modulated by adjusting the magnitudes of fields $\Omega$ and $\Omega'$, which allows our scheme to be directly applied in the twisting echo protocol proposed in Ref. \cite{Davis2016,Nolan2017} that is robust against detection noise.

\section{numerical investigation of squeezing dynamics}

Now we investigate the evolutions of the quantum state with constant drivings. Demonstrated by Husimi Q representation on the generalized Bloch spheres, the isotropic variance of the
initial CSS of spin $S$ is continuously
redistributed and reduced in a certain direction, indicating the spin
squeezing, as is shown in Fig.~\ref{fig:1}(d). The degree of spin squeezing is usually quantified by the
squeezing parameter $\xi ^{2}=4(\Delta S_{\perp })_{\mathrm{min}}^{2}/N_{s}$
\cite{Kitagawa1993}, where $(\Delta S_{\perp })_{%
	\mathrm{min}}^{2}$ is the minimum of the fluctuation $(\Delta S_{\perp
})^{2}=\langle S_{\perp }^{2}\rangle -\langle S_{\perp }\rangle ^{2}$ for
the spin component perpendicular to the mean spin direction. We compare the
squeezing parameters for free evolution under the three Hamiltonians $%
H_{1},H_{2},H_{3}$, and for the evolution of $H_{2}$ with constant drivings,
as shown in Fig.~\ref{fig:1}(e). It clearly shows that our scheme largely
improves the squeezing properties. Note that applying constant drivings on
Hamiltonians $H_{1}$ and $H_{3}$ would also lead to the improvement to the
OAT squeezing.

\begin{figure}[t]
	\includegraphics[width=\columnwidth]{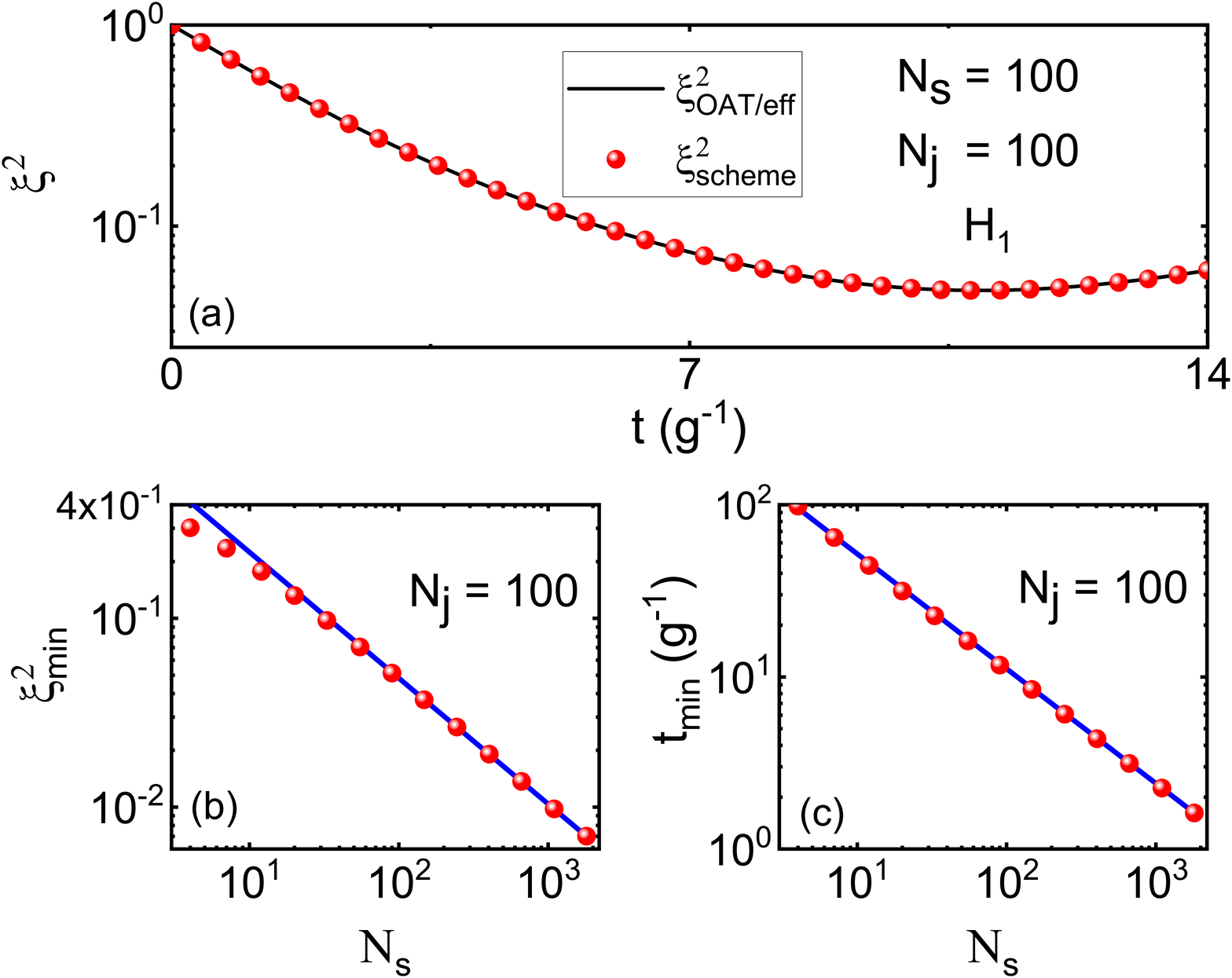}
	% Here is how to import EPS art
	\caption{The effective OAT spin squeezing under Hamiltonian $H_{1}$ with
		constant DC driving field. (a) Evolution of the spin squeezing parameter $\protect%
		\xi ^{2}$ of our scheme with constant drivings (red solid balls), compared
		with that of the corresponding effective OAT Hamiltonian (\protect\ref%
		{HOATeff}) (black solid curve). (b) and (c) demonstrate the optimal spin
		squeezing parameter $\protect\xi _{\mathrm{min}}^{2}$ and the corresponding
		squeezing time $t_{\mathrm{min}}$ as functions of particle number $N_{s}$
		(red solid balls). The blue solid lines are predicted by equation~(\protect
		\ref{xi&tmin}). The parameter $\Omega =50N_{j}g$ with particle numbers given
		in each subgraph.}
	\label{fig:2}
\end{figure}

According to the previous conclusion of OAT squeezing \cite{Jin2009} and
based on the obtained effective interaction coefficient $\chi _{\mathrm{eff}}
$, we derive the optimal spin squeezing parameter and the corresponding
squeezing time as
\begin{equation}
	\xi _{\mathrm{min}}^{2}\simeq \frac{1}{2}(\frac{N_{s}}{3})^{-\frac{2}{3}},%
	\text{ }t_{\mathrm{min}}\simeq \frac{2\times 3^{1/6}\Delta }{g^{2}N_{s}^{2/3}%
	}.  \label{xi&tmin}
\end{equation}%
The validity of effective OAT squeezing is further demonstrated in Fig.~\ref%
{fig:2}. The evolution of the squeezing parameter with constant drivings
agrees well with the corresponding effective OAT Hamiltonian (\ref{HOATeff}%
), as is shown in Fig.~\ref{fig:2}(a). The power-law scalings given in Eq.~(\ref{xi&tmin}) are also verified in Fig.~\ref{fig:2}(b) and (c).

Although Eq.~(\ref{Hlmgeff}) shows that both $S_{x}^{2}$ and $S_{y}^{2}$
exist, the tuning ranges of parameters $p$ and $q$ are not broad enough to
directly obtain the effective TAT interaction with the form $\propto
(S_{i}^{2}-S_{j}^{2})$. This imperfection can be overcome by adding an
additional continuous AC driving field, e.g., $H_{\mathrm{driv}}^{\mathrm{AC}%
}=A\cos (\omega t)S_{z}$, with $A$ and $\omega $ being the amplitude and
frequency of the driving field, respectively. The total driving terms
including $H_{\mathrm{driv}}^{\mathrm{DC}}$ and $H_{\mathrm{driv}}^{\mathrm{%
		AC}}$ become
\begin{equation}
	H_{\mathrm{driv}}=\Omega J_{z}+\Omega ^{\prime }S_{z}+A\cos (\omega t)S_{z},
	\label{Hdrivtot}
\end{equation}%
and the total Hamiltonian then becomes $H_{\mathrm{tot}}^{\prime }=H_{%
	\mathrm{int}}+H_{\mathrm{dirv}}$. Now we apply two transformations on the total
Hamiltonian, one is the same FNT as described by Eq. (\ref{FNT1}), the other
is $U_\mathrm{I}(t)=\exp [-i\int_{0}^{t} H_{\mathrm{driv}}^{\mathrm{AC}}(\tan)d\tau]$. Then the Hamiltonian can be
simplified as $H_{\mathrm{I}}\simeq[ (p+q)/2-(q-p)J_{0}(2A/\omega )/2] S_{x}^{2}+[ (p+q)/2+(q-p)J_{0}(2A/\omega )/2] S_{y}^{2}$, where $J_{n}(z)$ is the $n$-th Bessel function of the first kind. Therefore,
when $A/\omega $ is properly chosen, $H_\mathrm{I}$ will be a TAT
Hamiltonian under the condition $(p-2q)(2p-q)\geq 0$. For other cases, e.g.,
$p=q$, the continuous AC driving field can be applied along the $y$
direction with $H_{\mathrm{driv}}^{\mathrm{AC}}=A\cos (\omega t)S_{y}$,
which yields
\begin{equation}
	H_{\mathrm{eff}}^{\mathrm{TAT}}=\frac{\chi _{\mathrm{eff}}}{3}%
	(S_{x}^{2}-S_{y}^{2}),  \label{HTATeff}
\end{equation}%
where $\chi _{\mathrm{eff}}$ is the same as the OAT case (see Appendix \ref{App2} for detailed derivations).
\begin{figure}[t]
	\includegraphics[width=\columnwidth]{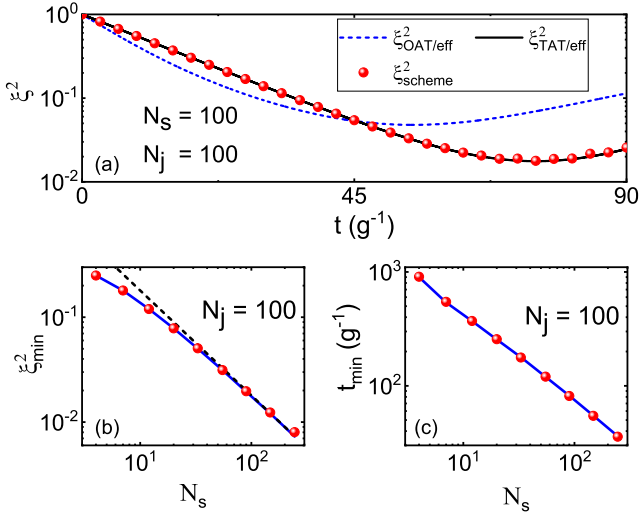}
	% Here is how to import EPS art
	\caption{The effective TAT spin squeezing under Hamiltonian $H_{2}$ with DC
		and AC driving fields. (a) Evolution of the spin squeezing parameter $%
		\protect\xi ^{2}$ for coupled spin systems with drivings (red solid balls)
		and for the effective TAT Hamiltonian (\protect\ref{HTATeff}) (black solid
		curve), compared with the OAT spin squeezing (blue dashed curve) governed by
		Hamiltonian (\protect\ref{HOATeff}). (b) and (c): The optimal spin squeezing
		$\protect\xi _{\mathrm{min}}^{2}$ and the optimal squeezing time $%
		t_{\mathrm{min}}$ as functions of the particle number $N_{s}$ for the TAT
		scheme with drivings (red solid balls), compared with the effective TAT
		results (blue solid curves). The black dashed line in (b) corresponds to $%
		\protect\xi _{\mathrm{min}}^{2}=1.8/N_{s}$. The parameter $\Delta/g =100$ and
		particle numbers are given in each subgraph.}
	\label{fig:3}
\end{figure}

To verify the
validity of our TAT scheme, we numerically study the evolution of spin
squeezing parameter of Hamiltonian (\ref{H}), coupled with the total driving
(\ref{Hdrivtot}), compared with the effective TAT Hamiltonian (\ref{HTATeff}%
). As is shown in Fig.~\ref{fig:3}(a), the evolution of spin squeezing
parameter under our scheme agrees well with that of the effective TAT
Hamiltonian dynamics, performing much better than the OAT spin squeezing. As
the effective interaction strength of TAT spin squeezing is obtained in Eq.~(%
\ref{HTATeff}) ($\chi _{\mathrm{eff}}/3$), the power-law scalings can be approximately obtained according to
the standard TAT squeezing \cite{Liu2011}
\begin{equation}
	\xi _{\mathrm{min}}^{2}\simeq \frac{1.8}{N_{s}},\text{ }t_{\mathrm{min}%
	}\simeq \frac{3\Delta \ln (4N_{s})}{g^{2}N_{s}}.  \label{xi&tminTAT}
\end{equation}%
They are also verified in Fig.~\ref{fig:3}(b) and (c). Therefore, adding both
DC and AC driving fields will finally transform the initial interaction (\ref{H}) into the TAT interaction, with squeezing degree up to the
Heisenberg-limited measurement precision.

\begin{figure}[t]
	\includegraphics[width=1\columnwidth]{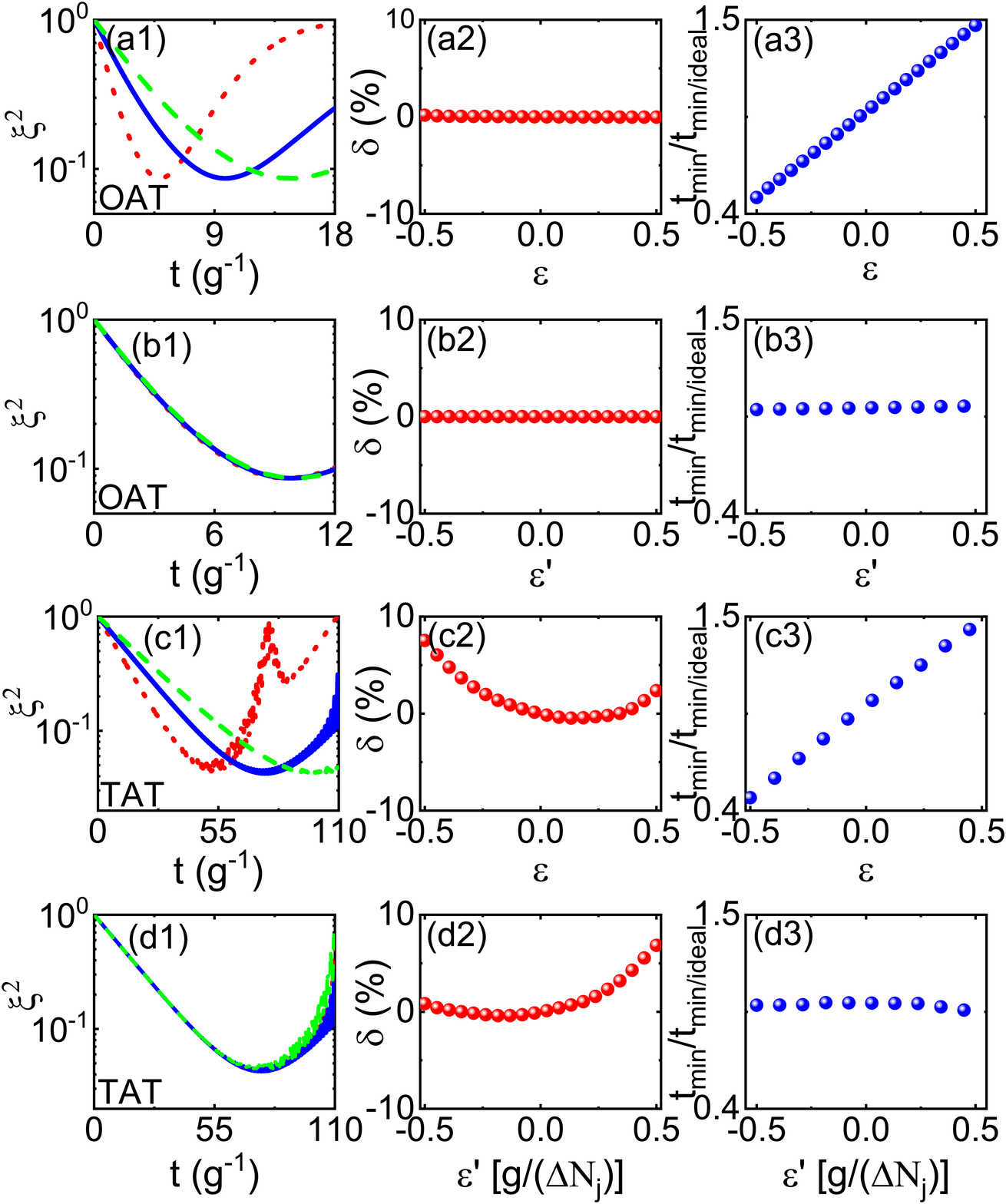}
	% Here is how to import EPS art
	\caption{The influence of parameter imperfections. The first row (a1-a3)
		show the effective OAT spin squeezing under different deviations of $\Omega
		$, denoted as $\protect\varepsilon $. The second row (b1-b3) are the results
		for the deviation of $\Omega ^{\prime }$, denoted as $\protect\varepsilon %
		^{\prime }$. The third (c1-c3) and fourth (d1-d3) rows are the corresponding
		results of the effective TAT spin squeezing under different $\protect%
		\varepsilon $ and $\protect\varepsilon ^{\prime }$, respectively. The
		deviations $\protect\varepsilon $, $\protect\varepsilon ^{\prime }$ and  relative error $\delta$ are
		defined in the main text. For the first column (a1-d1), the dotted red, solid blue, dashed green curves correspond to (a1) $\protect\varepsilon=-0.5,0,0.5$, (b1) $\protect\varepsilon^{\prime}=-1,0,1$, (c1) $\protect\varepsilon=-0.3,0,0.3$ and (d1) $\protect\varepsilon^{\prime}=-2\times10^{-5},0,2\times10^{-5}$, respectively. The parameters are $\Delta /g=50$ for OAT and $%
		\Delta /g=250$ for TAT spin squeezing, and the particle numbers are $%
		N_{s}=N_{j}=40$.}
	\label{fig:4}
\end{figure}

\section{discussion on parameter imperfections}

Considering the realistic experimental system, we investigate the influence
of parameter imperfections. On one hand, during the derivation of effective
interaction, the magnitudes of $\Omega $ and $\Omega ^{\prime }$ should
satisfy the constraint $f=0$ %\red{(\ref{constraint})}%
in order to eliminate the linear term in the effective Hamiltonian. Here we
investigate the influence of the deviation of $\Omega $ or $\Omega ^{\prime }
$ from their ideal values, keeping the value of the other one unchanged. The
deviations are quantified by the relative error, defined as $\varepsilon
=(\Omega -\Omega _{\mathrm{ideal}})/\Omega _{\mathrm{ideal}}$ for $\Omega $
and $\varepsilon ^{\prime }=(\Omega ^{\prime }-\Omega _{\mathrm{ideal}%
}^{\prime })/\Omega _{\mathrm{ideal}}^{\prime }$ for $\Omega ^{\prime }$,
respectively. As shown in Fig.~\ref{fig:4}(a1)-(d1) (the first column), the evolutions of the
spin squeezing parameters $\xi ^{2}$ deviate from the ideal cases to some
extent, but the squeezing is not much degraded. The optimal squeezing is insensitive to the deviations for OAT while slightly sensitive for TAT squeezing, as shown in Fig.~\ref{fig:4}(a2)-(d2) (the second column) with the relative error of optimal squeezing parameter $\delta=(\xi_\mathrm{min}^2-\xi_\mathrm{min0}^2)/\xi_\mathrm{min0}^2$. This is due to the fact that the effective TAT squeezing needs
the rotation of twisting axis, which is susceptible to the linear term in
the total Hamiltonian. Nevertheless, it still stays within the relative
error of $10\%$ for relatively large deviations, as shown in Fig.~\ref{fig:4}%
(c2)-(d2). Therefore, our scheme is overall robust to the deviation from the ideal combination of $\Omega$ and $\Omega'$. The optimal squeezing times $t_{%
	\mathrm{min}}$ increase linearly as the deviation of $\Omega $ increases,
but are not sensitive to the deviation of $\Omega ^{\prime }$. This is
because $\chi _{\mathrm{eff}}$
is in inverse proportional to $\Delta \propto (\Omega -\Omega ^{\prime })$,
and $\Omega $ is assumed to be much larger than $\Omega ^{\prime }$ in the
plots.

\begin{figure}[t]
	\includegraphics[width=1\columnwidth]{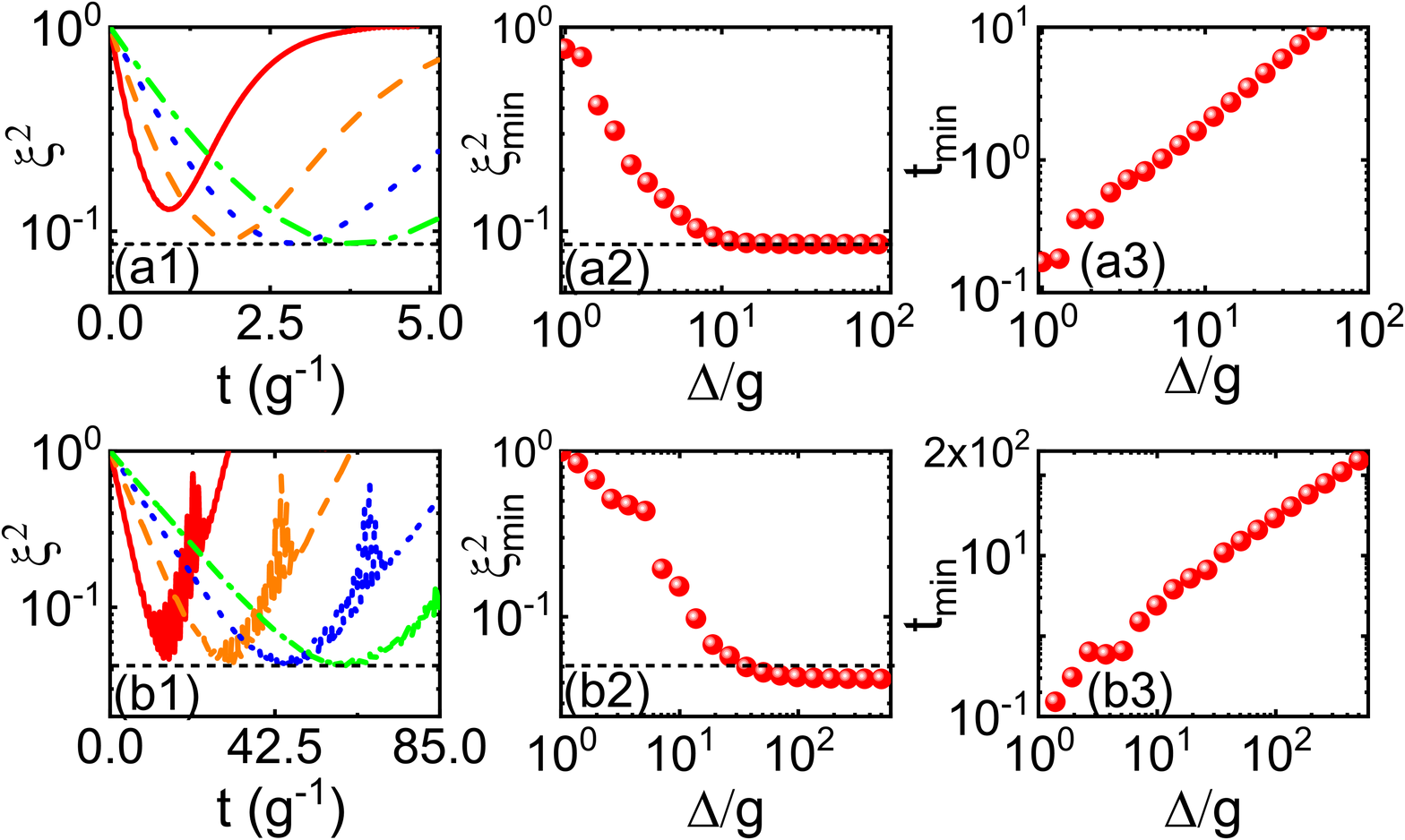}
	% Here is how to import EPS art
	\caption{Spin squeezing under different $\Delta $. The top panel (a1-a3)
		demonstrates the evolution of spin squeezing parameter $\protect\xi ^{2}$
		under different $\Delta $, the optimal spin squeezing $\protect\xi _{\mathrm{%
				min}}^{2}$ and the optimal squeezing time $t_{\mathrm{min}}$ as functions of
		$\Delta $, for the OAT spin squeezing.  The bottom panel (b1-b3) show the results for the TAT spin
		squeezing. For the first column (a1-b1), the solid red, dashed orange, dotted blue, dash-dotted green curves correspond to (a1) $\Delta/g=5,10,15,20$ and (a2) $\Delta/g=50,100,150,200$, respectively. The horizontal dashed lines denote the optimal spin squeezing achieved by a standard OAT 
		squeezing in (a1-a2) and a standard TAT one in (b1-b2). The particle numbers are $N_{s}=N_{j}=40$.}
	\label{fig:5}
\end{figure}

On the other hand, we investigate the
influence of $\Delta $ on the effective spin squeezing, as is
shown in Fig.~\ref{fig:5}. Overall, the best attainable squeezing $\xi _{%
	\mathrm{min}}^{2}$ becomes better with the increase of $\Delta $ [Fig.~\ref%
{fig:5}(a2) and (b2)], while the optimal squeezing time $t_{\mathrm{min}}$
also increases [Fig.~\ref{fig:5}(a3) and (b3)]. Specifically, given the
particle numbers, when the value of $\Delta $ is larger than a certain
value, the best attainable spin squeezing can achieve the optimal spin
squeezing for the corresponding effective OAT or TAT spin squeezing, as
shown in Fig.~\ref{fig:5}(a2) and (b2). Approximately, it requires $\Delta
/g\gg 10$ for OAT squeezing and $\Delta /g\gg 50$ for TAT squeezing, which
shows that the condition for realizing TAT squeezing is relatively stringent
than that of OAT, and the required squeezing time seems to be longer.
Nevertheless, the above discussions are limited by the particle numbers due
to the constrain of numerical computation resources. Since the optimal
squeezing time $t_{\mathrm{min}}$ scales as $\ln (4N_{s})/N_{s}$ for TAT
squeezing, for very large $N_{s}$, the required squeezing time will not need
to be too long.

\section{conclusion}

In summary, we have proposed a universal scheme to generate the effective
OAT and TAT spin squeezing in a broad category of coupled spin systems with
collective interaction Hamiltonian
$H_{\mathrm{int}}=\sum\nolimits_{\mu }g_{\mu }S_{\mu }J_{\mu }$,
by applying only constant and continuous drivings, which are simple to
implement. Both the best attainable spin squeezing and the
corresponding optimal time as functions of particle number are demonstrated
to satisfy the power-law scalings of OAT or TAT squeezing. In particular,
Heisenberg-limited measurement precision can be reached in such coupled spin
systems. Furthermore, our scheme is demonstrated to be tolerant to the
parameter imperfections of the driving fields. This work offers the
opportunity to realize Heisenberg-limited spin squeezing in a variety of
coupled spin systems that are common in realistic physical systems.

\begin{acknowledgments}
	
This work is supported by the Key-Area Research and Development Program of
Guangdong Province (Grant No.~2019B030330001), the National Natural Science
Foundation of China (NSFC) (Grant Nos. 12275145, 92050110, 91736106, 11674390, and
91836302), and the National Key R\&D Program of China (Grants No.
2018YFA0306504).

\end{acknowledgments}

\appendix

\section{Derivation of the effective squeezing Hamiltonian}

\label{App1}

Starting from the total Hamiltonian
\begin{equation}
  H_\mathrm{tot} = \Omega J_z + \Omega' S_z + \sum_\mu g_\mu S_\mu J_\mu,
\end{equation}
we perform the Fr\"ohlich-Nakajima transformation (FNT) to obtain the squeezing Hamiltonian.

To be specific, we introduce a unitary transformation
\begin{equation}
  U = e^S,
\end{equation}
where
\begin{equation}
  S = -i(\theta_{xy}S_xJ_y+\theta_{yx}S_yJ_x).
\end{equation}
The Hamiltonian is then transformed into
\begin{equation}\label{H'}
  \begin{split}
    H' &= e^{-S}H_\mathrm{tot}e^{S} \\
    &= H_\mathrm{tot}+\comm{H_\mathrm{tot}}{S} + \frac{1}{2}\comm{\comm{H_\mathrm{tot}}{S}}{S} + ...
  \end{split}
\end{equation}
We calculate the commutator $\comm{H_\mathrm{tot}}{S}$ as
\begin{widetext}
  \begin{equation}
    \begin{split}
      \comm{H_\mathrm{tot}}{S} &= \Omega\comm{J_z}{-i(\theta_{xy}S_xJ_y+\theta_{yx}S_yJ_x)} + \Omega'\comm{S_z}{-i(\theta_{xy}S_xJ_y+\theta_{yx}S_yJ_x)}\\
      &\quad+ \comm{g_xS_xJ_x+g_yS_yJ_y+g_zS_zJ_z}{-i(\theta_{xy}S_xJ_y+\theta_{yx}S_yJ_x)}\\
      &= -(\theta_{xy}\Omega+\theta_{yx}\Omega')S_xJ_x + (\theta_{yx}+\theta_{xy}\Omega')S_yJ_y + g_x\theta_{xy}S_x^2Jz - g_y\theta_{yx}S_y^2J_z\\
      &\quad+ g_x\theta_{yx}S_zJ_x^2 - g_y\theta_{xy}S_zJ_y^2 + g_z\theta_{xy}(-S_zS_xJ_x + S_yJ_yJ_z) + g_z\theta_{yx}(S_zS_yJ_y-S_xJ_xJ_z),
    \end{split}
  \end{equation}
\end{widetext}
and further obtain
\begin{widetext}
  \begin{equation}
    \begin{split}
      \comm{\comm{H_\mathrm{tot}}{S}}{S} = &-(\theta_{xy}\Omega+\theta_{yx}\Omega')\theta_{xy}S_x^2\comm{J_x}{-iJ_y} + (\theta_{yx}\Omega+\theta_{xy}\Omega')\theta_{xy}\comm{S_y}{-iS_x}J_y^2\\
      &- (\theta_{xy}\Omega+\theta_{yx}\Omega')\theta_{yx}\comm{S_x}{-iS_y}J_x^2 + (\theta_{yx}\Omega+\theta_{xy}\Omega')\theta_{yx}S_y^2\comm{J_y}{-iJ_x} + ...\\
      = &-(\theta_{xy}\Omega+\theta_{yx}\Omega')\theta_{xy}S_x^2J_z - (\theta_{yx}\Omega+\theta_{xy}\Omega')\theta_{xy}S_zJ_y^2\\
      &- (\theta_{xy}\Omega+\theta_{yx}\Omega')\theta_{yx}S_zJ_x^2 - (\theta_{yx}\Omega+\theta_{xy}\Omega')\theta_{yx}S_y^2J_z + ...
    \end{split}
  \end{equation}
\end{widetext}
Replace them into \eqref{H'}, we get
\begin{widetext}
  \begin{equation}
    \begin{split}
      H' &= \Omega J_z + \Omega' S_z + (g_x-\theta_{xy}\Omega-\theta_{yx}\Omega')S_xJ_x + (g_y+\theta_{yx}\Omega+\theta_{xy}\Omega')S_yJ_y + g_zS_zJ_z\\
      &\quad+ \left[g_x-\frac{1}{2}(\theta_{xy}\Omega+\theta_{yx}\Omega')\right]\theta_{xy}S_x^2J_z - \left[g_y+\frac{1}{2}(\theta_{yx}\Omega+\theta_{xy}\Omega')\theta_{yx}\right]S_y^2J_z\\
      &\quad+ \left[g_x-\frac{1}{2}(\theta_{xy}\Omega+\theta_{yx}\Omega')\right]\theta_{yx}S_zJ_x^2 - \left[g_y+\frac{1}{2}(\theta_{yx}\Omega+\theta_{xy}\Omega')\theta_{xy}\right]S_zJ_y^2\\
      &+ g_z\theta_{xy}(-S_zS_xJ_x+S_yJ_yJ_z) + g_z\theta_{yx}(S_zS_yJ_y-S_xJ_xJ_z) + ...
    \end{split}
  \end{equation}
\end{widetext}

In the spirit of FNT, we require
\begin{equation}
  \begin{split}
    &g_x - \theta_{xy}\Omega - \theta_{yx}\Omega' = 0,\\
    &g_y + \theta_{yx}\Omega + \theta_{xy}\Omega' = 0,
  \end{split}
\end{equation}
which leads to
\begin{equation}
  \theta_{xy} = \frac{g_x\Omega+g_y\Omega'}{\Omega^2-\Omega'^2},\theta_{yx}=-\frac{g_y\Omega+g_x\Omega'}{\Omega^2-\Omega'^2}.
\end{equation}
The expression of $H'$ now becomes
\begin{widetext}
  \begin{equation}
    \begin{split}
      H' &= \Omega J_z + \Omega' S_z + g_zS_zJ_z + \frac{g_x^2\Omega+g_xg_y\Omega'}{2(\Omega^2-\Omega'^2)}S_x^2J_z + \frac{g_y^2\Omega+g_xg_y\Omega'}{2(\Omega^2-\Omega'^2)}S_y^2J_z - \frac{g_x^2\Omega'+g_xg_y\Omega}{2(\Omega^2-\Omega'^2)}S_zJ_x^2\\
      &\quad-\frac{g_y^2\Omega'+g_xg_y\Omega}{2(\Omega^2-\Omega'^2)}S_zJ_y^2 + \frac{g_xg_z\Omega+g_yg_z\Omega'}{\Omega^2-\Omega'^2}(-S_zS_xJ_x+S_yJ_yJ_z) - \frac{g_yg_z\Omega+g_xg_z\Omega'}{\Omega^2-\Omega'^2}(S_zS_yJ_y-S_xJ_xJ_z) + ...
    \end{split}
  \end{equation}
\end{widetext}
In order to ignore higher order terms safely, the derivations above require
\begin{equation}\label{req}
  \frac{gN_\mathrm{s}}{\abs{\Omega-\Omega'}}\ll 1 \quad \& \quad \frac{gN_\mathrm{j}}{\abs{\Omega-\Omega'}}\ll 1.
\end{equation}

When focusing on the evolution of subsystem $\mathbf{S}$, we choose the initial state of subsystem $\mathbf{J}$ as the coherent spin state along $+z$ axis, i.e., $\ket{\psi(t=0)}_\mathrm{J}=\ket{z}$, and assume it is almost unchanged during evolution. Then we can replace the operators of subsystem $\mathbf{J}$ with their expected values:
\begin{equation}
  \begin{split}
    &\expval{J_z} = \frac{N_j}{2},\quad \expval{J_x^2} = \expval{J_y^2} = \frac{N_j}{4},\\
    &\expval{J_x} = \expval{J_y} = \expval{J_zJ_y} = \expval{J_zJ_x} = 0.
  \end{split}
\end{equation}
After that, we finally obtain the effective Hamiltonian for the subsystem $\mathbf{S}$:
\begin{equation}
  H_\mathrm{eff} = fS_z + pS_x^2 + qS_y^2,
\end{equation}
where
\begin{equation}
  \begin{split}
    &f = \Omega'+\frac{1}{2}g_zN_j-\frac{N_j}{8(\Omega^2-\Omega'^2)}\left[(g_x^2+g_y^2)\Omega'+2g_xg_y\Omega\right],\\
    &p = \frac{N_j(g_x^2\Omega+g_xg_y\Omega')}{4(\Omega^2-\Omega'^2)},\quad q = \frac{N_j(g_y^2\Omega+g_xg_y\Omega')}{4(\Omega^2-\Omega'^2)}.
  \end{split}
\end{equation}

For three typical interactions $H_1=gS_xJ_x,H_2=g(S_xJ_x+S_yJ_y+S_zJ_z),H_3=g(S_xJ_x+S_yJ_y-2S_zJ_z)$, the corresponding effective Hamiltonians are
\begin{equation}
  \begin{split}
    &H_\mathrm{1eff} = \left[\Omega'-\frac{g^2\Omega'N_j}{8(\Omega^2-\Omega'^2)}\right]S_z + \chi_\mathrm{1eff}S_x^2,\\
    &H_\mathrm{2eff} = \left[\Omega'+\frac{1}{2}gN_j-\frac{g^2\Omega'N_j}{4(\Omega^2-\Omega'^2)}\right]S_z + \chi_\mathrm{2eff}S_z^2,\\
    &H_\mathrm{3eff} = \left[\Omega'-gN_j-\frac{g^2\Omega'N_j}{4(\Omega^2-\Omega'^2)}\right]S_z + \chi_\mathrm{3eff}S_z^2,
  \end{split}
\end{equation}
with effecitve interaction strength $\chi_\mathrm{1eff} = g^2\Omega N_j/[4(\Omega^2-\Omega'^2)]$, $\chi_\mathrm{2eff} = \chi_\mathrm{3eff} = -g^2N_j/[4(\Omega-\Omega')]$. For $H_2$ and $H_3$, we have used the identity $S_x^2+S_y^2=S^2-S_z^2=s(s+1)-S_z^2$ and ignored the constant terms. The linear term $\propto S_z$ can be eliminated by choosing appropriate magnitudes of driving fields to make $f=0$, then all three effective Hamiltonians are reduced to pure OAT Hamiltonians $H_\mathrm{eff}^\mathrm{OAT} = \chi_\mathrm{eff}S_\mu^2$.

\section{Generation of the effecitve TAT Hamiltonian with continuous driving}

\label{App2}

An effective Hamiltonian with the form $pS_x^2 + qS_y^2$ can be obtained by adding constant DC driving fields in coupled spin systems, as shown in appendix \ref{App1}. If $p=-q$ or $p=2q$ or $p=q/2$ is satisfied, the effective TAT interaction $\propto (S_i^2-S_j^2)(i,j=x,y,z)$ can be obtained. Unfortunately, considering the condition of Eq. \eqref{req}, the tuning ranges of parameters $p$ and $q$ are not broad enough to directly obtain the effective TAT interaction. This imperfection can be overcome by adding an additional driving field. As suggested in Refs. \cite{Liu2011,Huang2015}, we could transform an OAT Hamiltonian into an effective TAT Hamiltonian by using a pulsed or continuous driving. Here we provide a general scheme to transform any Hamiltonian of the form $pS_x^2 + qS_y^2$ into a pure effective TAT Hamiltonian with continuous driving.

If $(p-2q)(2p-q)\geq0$, we add a continuous AC field along $z$ axis and get
\begin{equation}
  H = pS_x^2 + qS_y^2 + AS_z\cos\omega t.
\end{equation}
In the interaction picture difined by
\begin{equation}
  \begin{split}
    &\ket{\psi_\mathrm{I}(t)} = U_\mathrm{I}(t)\ket{\psi(t)},\\
    &H_\mathrm{I} = U_\mathrm{I}^\dagger(t)H_0U_\mathrm{I}(t),
  \end{split}
\end{equation}
where
\begin{equation}
  \begin{split}
    H_0 &= pS_x^2 + qS_y^2,\\
    U_\mathrm{I} &= \exp(-i\int_0^tAS_z\cos\omega\tau d\tau)\\
    &= \exp(-i\frac{A}{\omega}S_z\sin\omega t),
  \end{split}
\end{equation}
we can obtain
\begin{widetext}
  \begin{equation}
    \begin{split}
      H_\mathrm{I} &= \left[\frac{p+q}{2}-\frac{q-p}{2}\cos(\frac{2A}{\omega}\sin\omega t)\right] S_x^2 + \left[\frac{p+q}{2}+\frac{q-p}{2}\cos(\frac{2A}{\omega}\sin\omega t)\right] S_y^2\\
      &\quad +\frac{q-p}{2}\sin(\frac{2A}{\omega}\sin\omega t)(S_xS_y+S_yS_x),
    \end{split}
  \end{equation}
\end{widetext}
where we have used $e^{i\phi S_z}S_xe^{-i\phi S_z} = S_x\cos\phi-S_y\sin\phi$ and $e^{i\phi S_z}S_ye^{-i\phi S_z} = S_y\cos\phi+S_x\sin\phi$.

Now we apply the Jacobi-Anger expansion $e^{iz\sin\theta}=\sum_{n=-\infty}^\infty J_n(z)e^{in\theta}$, where $J_n(z)$ is the $n$th Bessel function of the first kind, and only keep the zero-order term with $n=0$ (rotating wave approximation), the Hamiltonian becomes
\begin{widetext}
  \begin{equation}\label{HI}
    H_\mathrm{I} \approx \left[\frac{p+q}{2}-\frac{q-p}{2}J_0(\frac{2A}{\omega})\right]S_x^2 + \left[\frac{p+q}{2}+\frac{q-p}{2}J_0(\frac{2A}{\omega})\right]S_y^2. 
  \end{equation}
\end{widetext}
This approximation requires $\omega\ll N_\mathrm{s}(p+q)$.

The Hamiltonian $H_\mathrm{I}$ can be rewritten by adding a constant term $-[(p+q)/2 - (q-p)J_0(2A/\omega)/2]S^2$ or $-[(p+q)/2 + (q-p)J_0(2A/\omega)/2]S^2$, which leads to
\begin{widetext}
  \begin{equation}
    \begin{split}
      &H_\mathrm{I}^{(1)} = (q-p)J_0(\frac{2A}{\omega})S_y^2 - \left[ \frac{p+q}{2}-\frac{q-p}{2}J_0(\frac{2A}{\omega}) \right]S_z^2,\\
      &H_\mathrm{I}^{(2)} = (p-q)J_0(\frac{2A}{\omega})S_x^2 - \left[ \frac{p+q}{2}+\frac{q-p}{2}J_0(\frac{2A}{\omega}) \right]S_z^2.
    \end{split}
  \end{equation}
\end{widetext}
The effective TAT Hamiltonian is obtained when setting $J_0(2A/\omega)=\pm(p+q)/[3(q-p)]$ ("$+$" for (1) and "$-$" for (2)), that is,
\begin{equation}
  \begin{split}
    &H_\mathrm{eff}^{(1)} = \frac{p+q}{3}(S_y^2-S_z^2),\\
    &H_\mathrm{eff}^{(2)} = \frac{p+q}{3}(S_x^2-S_z^2).\\
  \end{split}
\end{equation}

If $(p-2q)(2p-q)<0$, we may add a continuous AC field along $y$ axis. The Hamiltonian is equivalent to
\begin{equation}
  H' = -qS_z^2 + (p-q)S_x^2 + AS_y\cos\omega t.
\end{equation}
With similar analysis, we find the effective TAT Hamiltonian is obtained when $J_0(2A/\omega)=\pm(p-2q)/3p$, and the result is
\begin{equation}
  \begin{split}
    &{H_\mathrm{eff}'}^{(1)} = \frac{p-2q}{3}(S_x^2-S_y^2),\\
    &{H_\mathrm{eff}'}^{(2)} = \frac{p-2q}{3}(S_z^2-S_y^2).\\
  \end{split}
\end{equation}

\bibliography{ESSCSSBIB}

\end{document}